\documentclass[reprint, aps, pra, floatfix]{revtex4-2}
\usepackage[dvips]{graphicx}
\usepackage[english]{babel}
\usepackage{amsmath}
\usepackage{amssymb}
\usepackage{bm}
\usepackage{subfigure}
\usepackage[colorlinks, citecolor = black, linkcolor = black, urlcolor = blue]{hyperref}

\newcommand{\bk}{\mathbf{k}}
\newcommand{\bq}{\mathbf{q}}
\newcommand{\nn}{\nonumber}
\newcommand{\p}{\partial}
\newcommand{\ak}{a_\mathbf{k}}
\newcommand{\amk}{a_{-\mathbf{k}}}

\newcommand{\hh}{\hat{H}}
\newcommand{\tr}{\mathrm{Tr}}

\newcommand{\bn}{\mathbf{n}}

\def\bra#1{\langle#1|}
\def\ket#1{|#1\rangle}
\def\braket#1{\langle#1\rangle}
\def\ketbra#1#2{|#1\rangle\langle#2|}

\def\bf#1{\mathbf{#1}}

\begin{document}
\title{Complexity of Bose-Einstein condensates at finite temperature}
\author{Chang-Yan Wang}
\email{changyanwang@tsinghua.edu.cn}
\affiliation{Institute for Advanced Study, Tsinghua University, Beijing 100084, China}
\begin{abstract}
  We investigate the geometric quantum complexity of Bose-Einstein condensate (BEC) at finite temperature. Specifically, we use the Bures and Sj\"oqvist metrics---generalizations of the Fubini-Study metric for mixed quantum states, as well as the Nielsen geometric complexity approach based on purification of mixed states. Starting from the Bogoliubov Hamiltonian of BEC, which exhibits an $SU(1,1)$ symmetry, we explicitly derive and compare the complexities arising from these three distinct measures. For the Bures and Sj\"oqvist metrics, analytical and numerical evaluations of the corresponding geodesics are provided, revealing characteristic scaling behaviors with respect to temperature. In the Nielsen complexity approach, we rigorously handle the gauge freedoms associated with mixed state purification and non-uniqueness unitary operations, demonstrating that the resulting complexity aligns precisely with the Bures metric. Our work provides a comparative study of the geometric complexity of finite-temperature Bose-Einstein condensates, revealing its intimate connections to symmetry structures and temperature effects in BEC systems.
\end{abstract}
\maketitle

\section{Introduction}
Geometric quantum complexity, originally developed in quantum information theory, quantifies the minimal resources required to implement a quantum operation or prepare a quantum state from a geometric perspective~\cite{nielsen_geometric_2005, nielsen_quantum_2006, chapman_definition_2018}. Recent advances have extended this concept across various areas of quantum physics, including quantum field theory~\cite{jefferson_circuit_2017, chapman_complexity_2017, susskind_computational_2016, caceres_complexity_2020, ruan_purification_2021}, quantum gravity~\cite{brown_complexity_2016, carmi_time_2017, couch_holographic_2018, caputa_quantum_2019}, and condensed matter physics~\cite{liu_circuit_2020, rocajerat_circuit_2023, sood_circuit_2022, suzuki_quantum_2025, suzuki_quantum_2025, znidaric_complexity_2008, lv_building_2024}, where complexity serves as a fundamental measure of quantum evolution and entanglement \cite{eisert_entangling_2021}.

For pure states, geometric quantum complexity is often defined as the length of optimal geodesics on either the manifold of unitary operations, as introduced by Nielsen and collaborators~\cite{nielsen_geometric_2005,nielsen_quantum_2006}, or alternatively, the manifold of quantum states equipped with the Fubini-Study metric~\cite{chapman_definition_2018, brown_complexity_2019}. However, realistic quantum systems, particularly those at finite temperature or interacting with an environment, are described by mixed states. Extending the notion of complexity to mixed states is nontrivial due to the lack of a unique unitary orbit, the existence of multiple purifications, and the variety of possible metrics that define distances between mixed states~\cite{bengtsson_geometry_2006,alsing_comparing_2023}.

In this work, we investigate the geometric complexity of Bose-Einstein condensates (BECs) at finite temperature. BECs are an ideal platform for this study, as they form macroscopic quantum states that exhibit high coherence at low temperatures but transition to mixed states as thermal excitations populate higher modes with increasing temperature~\cite{anderson_observation_1995, davis_boseeinstein_1995, pethick_bose_2008, pitaevskii_boseeinstein_2016}. Furthermore, the Bogoliubov Hamiltonian governing BECs possesses an $SU(1,1)$ symmetry~\cite{wang_uhlmann_2024, chen_manybody_2020, cheng_manybody_2021, lv_su_2020, lyu_geometrizing_2020, zhang_quantum_2022, zhang_periodically_2020}, which constrains the density matrices of these mixed states to a three-dimensional cylindrical manifold—a natural geometric setting for analyzing complexity. Understanding the complexity of such states is of both fundamental and practical interest: fundamentally, it quantifies the minimal effort required to transform a mixed macroscopic quantum state~\cite{polkovnikov_colloquium_2011}; practically, it may inform the design of quantum simulators and optimize resources for creating or maintaining quantum states in cold-atom experiments~\cite{bloch_manybody_2008, gross_quantum_2017}.

To capture complexity in mixed states, we apply three geometric approaches: (i) the Bures metric \cite{dittmann_explicit_1999, bures_extension_1969, uhlmann_metric_1992, hubner_explicit_1992, hubner_computation_1993}, derived from Uhlmann fidelity~\cite{uhlmann_transition_1976}, which measures the distance between density matrices; (ii) the Sjöqvist metric~\cite{sjoqvist_geometric_2000, sjoqvist_geometry_2020}, which reflects distances between sets of rays from the spectral decomposition of the density matrix; (iii) Nielsen's circuit complexity, formulated through purification of the density matrix~\cite{nielsen_geometric_2005}.

For the Bures and Sjöqvist metrics, we explicitly compute geodesics within the cylindrical manifold of density matrices and reveal scaling behaviors with temperature. For Nielsen complexity, we carefully handle the gauge freedoms associated with purification and unitary operations, demonstrating that by minimizing over these freedoms, the resulting complexity coincides with the Bures metric. Our study provides a comparative analysis of geometric complexity in finite-temperature BECs, highlighting the interplay between symmetry and thermal effects.

This paper is organized as follows. In Sec.~\ref{sec:ham}, we introduce the BEC Hamiltonian and discuss its $SU(1,1)$ symmetry, defining the relevant mixed coherent states. In Sec.~\ref{sec:bures}, we derive and analyze complexity using the Bures metric. Sec.~\ref{sec:sjoqvist} presents a similar analysis using the Sjöqvist metric. In Sec.~\ref{sec:nielsen}, we formulate Nielsen complexity through purification, explicitly addressing gauge freedoms. We conclude with a summary and discussion in Sec.~\ref{sec:conclusion}.

\section{Mixed BEC coherent states} \label{sec:ham}
In this section, we briefly review the Bogoliubov Hamiltonian for BEC, and its underlying $SU(1,1)$ symmetry. By using the properties of $SU(1,1)$ coherent states, it can be shown that the mixed states of BEC at finite temperature are parameterized by a three-dimensional cylindrical manifold, which provides a natural setting to study the geometric complexity of these states.

We consider a weakly interacting bosonic gas, whose Hamiltonian in momentum space is given by
\begin{align}
  \hat{H} = \sum_\bk \epsilon_\bk a_\bk^\dag a_\bk + \frac{g}{2V} \sum_{\bk,\bk',\bq} a_{\bk + \bq}^\dag a_{\bk' - \bq}^\dag a_{\bk'} a_{\bk},
\end{align}
where $a_\bk$ ($a_\bk^\dag$) are the annihilation (creation) operators for bosons with momentum $\bk$. The kinetic energy is given by $\epsilon_\bk = \bk^2/2m - \mu$, where $m$ is the boson mass and $\mu$ is the chemical potential. The interaction strength is characterized by $g = 4\pi \hbar^2 a_s/m$, with $a_s$ being the $s$-wave scattering length, which can be tuned through Feshbach resonance \cite{chin_feshbach_2010}. In the following, we set $\hbar = 1$ for simplicity.

At sufficiently low temperatures, Bose-Einstein condensation (BEC) occurs, and the majority of bosons occupy the lowest energy state with $\bk=0$. By applying the Bogoliubov approximation, we expand the interaction terms around the condensate and keep terms up to second order in boson operators. This leads to the Bogoliubov Hamiltonian $\hh_{\mathrm{Bog.}} = \sum_{\bk \neq \mathbf{0}} \hh_\bk + \text{const.}$,
where the Hamiltonian for each nonzero momentum mode takes the form:
\begin{align}
  \hh_\bk = \sum_{i=0}^2 \xi_i(\bk) K_i.\label{ham}
\end{align}

The coefficients are defined as $\xi_0(\bk) = 2(\epsilon_\bk + g|\Psi_0|^2)$, $\xi_1(\bk) = 2\text{Re}(g\Psi_0^2)$, and $\xi_2(\bk) = -2\text{Im}(g\Psi_0^2)$, with $\Psi_0 = \sqrt{N_0/V} e^{i\theta}$ the condensate wavefunction, $N_0$ the condensate particle number, and $V$ the system volume. The operators $K_i$ in Eq. (\ref{ham}) are
\begin{align} \label{generators}
  K_0 &= \frac{1}{2}(a_\bk^\dag a_\bk + a_{-\bk} a_{-\bk}^\dag),  \nn\\
  K_1 &= \frac{1}{2}(a_\bk^\dag a_{-\bk}^\dag + a_\bk a_{-\bk}),  \nn\\
  K_2 &= \frac{1}{2i}(a_\bk^\dag a_{-\bk}^\dag - a_\bk a_{-\bk}), 
\end{align}
which form the generators of the $\mathfrak{su}(1,1)$ Lie algebra \cite{lyu_geometrizing_2020, puri_mathematical_2001}. They satisfy the commutation relations
\begin{align}
  [K_0, K_1] = i K_2, \, [K_2, K_0] = i K_1, \, [K_1, K_2] = -i K_0. \label{commu}
\end{align}
Since the Bogoliubov Hamiltonian $\hh_{\mathrm{Bog.}}$ is decoupled for different momenta, we focus on the single-mode Hamiltonian $\hh_\bk$ in Eq.\,(\ref{ham}).

To diagonalize $\hh_\bk$, we introduce the Bogoliubov transformation:
\begin{align}
  \hat{\alpha}_\bk &= u_\bk a_\bk + v_\bk a_{-\bk}^\dag, \quad
  \hat{\alpha}_{-\bk} = u_\bk a_{-\bk} + v_\bk a_\bk^\dag,
  \label{eq-bogo}
\end{align}
where the coefficients satisfy the normalization condition:
\begin{align} \label{sp}
  |u_\bk|^2 - |v_\bk|^2 = 1.
\end{align}
In terms of these quasi-particle operators, the Hamiltonian takes the diagonal form $\hh_\bk = \varepsilon_\bk(\hat{\alpha}_\bk^\dag \hat{\alpha}_\bk + \hat{\alpha}_{-\bk}^\dag \hat{\alpha}_{-\bk}) + \text{const.}$, with the quasi-particle dispersion relation $\varepsilon_\bk = \frac{1}{2} \sqrt{\xi_0(\bk)^2 - \xi_1(\bk)^2 - \xi_2(\bk)^2}$.

The ground state of $\hh_\bk$ is defined by $\hat{\alpha}_\bk |G\rangle = \hat{\alpha}_{-\bk} |G\rangle = 0$. This state is a generalized coherent state of the $SU(1,1)$ group \cite{perelomov_generalized_1986, zhai_ultracold_2021}:
\begin{align}
  |G\rangle = \frac{1}{\sqrt{1 - |z|^2}} e^{-z a_\bk^\dag a_{-\bk}^\dag} |0\rangle,
\end{align}
where $z = v_\bk / u_\bk$, and $|0\rangle$ is the bare vacuum. From Eq. (\ref{sp}), it follows that $|z| < 1$, ensuring $1 - |z|^2 > 0$. Thus, the ground states for all possible $\hh_\bk$ are parameterized by $z$ defined on a two-dimensional unit disk, which is actually the Poincare disk \cite{lyu_geometrizing_2020, perelomov_generalized_1986}.

Alternatively, the ground state can be expressed using the unitary displacement operator $D(\zeta)$
\begin{align}
  |G\rangle = D(\zeta) |0\rangle, \quad D(\zeta) = e^{-\zeta a_\bk^\dag a_{-\bk}^\dag + \zeta^* a_\bk a_{-\bk}}. \label{d_op}
\end{align}
By parameterizing the Poincaré disk coordinate as $z = \tanh\frac{r}{2} e^{i\theta}$, the displacement parameter takes the form $\zeta = \frac{r}{2}e^{i\theta}$ \cite{hasebe_sp_2020}. This enables an Euler decomposition of $D(\zeta)$ as \cite{hasebe_sp_2020}
\begin{align} \label{euler}
  D(\zeta) = e^{i\theta K_0} e^{-i r K_2} e^{-i \theta K_0},
\end{align}
which will be useful in later calculations.

The excited states are constructed using quasi-particle creation operators,
\begin{align}
  |\Psi_\mathbf{n}\rangle = \frac{1}{\sqrt{n_1! n_2!}} (\hat{\alpha}_\bk^\dag)^{n_1} (\hat{\alpha}_{-\bk}^\dag)^{n_2} |G\rangle, \label{psi_n}
\end{align}
where $\mathbf{n} = (n_1, n_2)$ denotes the number of $\bk$ and $-\bk$ quasi-particles. Noting that the Bogoliubov transformation satisfies $D(\zeta)^\dag \hat{\alpha}_{\pm\bk}^\dag D(\zeta) = a_{\pm\bk}^\dag$ \cite{puri_mathematical_2001, perelomov_generalized_1986}, we can rewrite the excited states as
\begin{align}
  |\Psi_\mathbf{n}\rangle = D(\zeta) \frac{1}{\sqrt{n_1! n_2!}} (a_\bk^\dag)^{n_1} (a_{-\bk}^\dag)^{n_2} |0\rangle = D(\zeta) |\mathbf{n}\rangle, \label{dn}
\end{align}
with $|\mathbf{n}\rangle = (n_1! n_2!)^{-1/2} (a_\bk^\dag)^{n_1} (a_{-\bk}^\dag)^{n_2} |0\rangle$. The eigenenergy of $|\Psi_\mathbf{n}\rangle$ is $E_\mathbf{n} = \varepsilon_\bk (n_1 + n_2)$, where we have shifted the ground-state energy to zero for simplicity, which does not affect the complexity calculations below.

At finite temperature, the BEC state at momenta $\pm\bk$ is described by a density matrix,
\begin{align}
  \rho = \sum_\mathbf{n} \lambda_\mathbf{n} |\Psi_\mathbf{n}\rangle \langle \Psi_\mathbf{n}|,
\end{align}
where $\lambda_\mathbf{n} = e^{-\beta \varepsilon_\bk (n_1 + n_2)} / \mathcal{Z}$, $\beta = 1/T$, and $\mathcal{Z} = \sum_\mathbf{n} e^{-\beta \varepsilon_\bk (n_1 + n_2)}$ is the partition function. Here, we set the Boltzmann constant as $k_B = 1$ for simplicity. Thus, $\rho$ is parameterized by $(\beta, r, \theta)$, forming a three-dimensional cylindrical manifold. In the following sections, we explore the geometric complexity of evolving a reference state $\rho_R$ to a target state $\rho_T$. For simplicity, we assume the quasi-particle energy $\varepsilon_\bk$ remains constant during this evolution, dropping the $\bk$ subscript and denoting it as $\varepsilon$. As we will show, this complexity corresponds to the geodesic length connecting these states on the cylinder, computed using different metrics.

\section{Complexity: Bures metric}\label{sec:bures}
One way to quantify the geometric complexity of pure quantum states is through the Fubini-Study metric, which measures distances between states in Hilbert space \cite{chapman_definition_2018}. A natural generalization of the approach to mixed states is to replace the Fubini-Study metric with a metric that measures the distance between mixed states, which can be done by the so-called Bures metric \cite{dittmann_explicit_1999, bures_extension_1969, uhlmann_metric_1992, hubner_explicit_1992, hubner_computation_1993}.  In this section, we employ the Bures metric to explore the geometric complexity of mixed coherent states in a BEC at finite temperature.

Given a Hilbert space $\mathcal{H}$ parameterized by coordinates $x^1, x^2, \dots, x^{\dim \mathcal{H}}$, the Fubini-Study metric measures the infinitesimal distance between a state $\ket{\psi} \in \mathcal{H}$ and its neighboring state $\ket{\psi + d\psi}$ as \cite{provost_riemannian_1980} 
\begin{align}\label{fs}
    ds_{\mathrm{FS}}^2 = 2\left(1 - |\braket{\psi | \psi + d\psi}|\right).
\end{align}
The complexity between two pure states $\ket{\psi_1}$ and $\ket{\psi_2}$ is then defined as the length of the shortest path, or geodesic, connecting them \cite{chapman_definition_2018}
\begin{align}
    \mathcal{C} = \int_{\Gamma_{\mathrm{min}}} ds_{\mathrm{FS}}.
\end{align}

For mixed states, we work in the space $\mathcal{M}$ of full-rank density matrices, parameterized by $x^1, x^2, \ldots, x^{\dim \mathcal{M}}$. The pure-state overlap $|\braket{\psi_1 | \psi_2}|$ is generalized to the Uhlmann fidelity between mixed states $\rho_1$ and $\rho_2$ \cite{uhlmann_transition_1976}
\begin{align}
    F(\rho_1, \rho_2) = \mathrm{Tr} \left( \sqrt{\sqrt{\rho_1} \rho_2 \sqrt{\rho_1}} \right).
\end{align}
This fidelity reduces to the overlap when both states are pure, making it a suitable generalization to mixed states. The Bures metric defines the infinitesimal distance between $\rho$ and $\rho + d\rho$ as \cite{bures_extension_1969, uhlmann_metric_1992}
\begin{align}
    ds_B^2 = 2 \left[ 1 - F(\rho, \rho + d\rho) \right].
\end{align}

The complexity between two mixed states $\rho_1$ and $\rho_2$ is thus the geodesic length in this metric. For a density matrix $\rho = \sum_i \lambda_i |u_i\rangle \langle u_i|$ in its diagonal basis, the Bures metric takes the form $ds_B^2 = g_{\mu\nu}^B dx^\mu dx^\nu$, where the metric tensor is \cite{hubner_explicit_1992, hubner_computation_1993}
\begin{align}\label{g_bures}
    g_{\mu\nu}^B = \frac{1}{2} \sum_{i,j} \frac{\langle u_i | \partial_\mu \rho | u_j \rangle \langle u_j | \partial_\nu \rho | u_i \rangle}{\lambda_i + \lambda_j}.
\end{align}
(See Appendix \ref{app:bures} for a detailed derivation.) Noting that $\partial_\mu \rho = \sum_i \partial_\mu \lambda_i |u_i \rangle \langle u_i| + \lambda_i | \partial_\mu u_i \rangle \langle u_i| + \lambda_i |u_i \rangle \langle \partial_\mu u_i |$, and using the fact that $\langle \partial_\mu u_i | u_j \rangle = - \langle u_i | \partial_\mu u_j \rangle$ (from $\partial_\mu\braket{u_i|u_j} = 0$), we have
\begin{align}
    g_{\mu\nu}^B = \sum_i \frac{\partial_\mu \lambda_i \partial_\nu \lambda_i}{4 \lambda_i} + \frac{1}{2} \sum_{i \neq j} \frac{(\lambda_i - \lambda_j)^2}{\lambda_i + \lambda_j} |\langle u_i | \partial_\mu u_j \rangle|^2.
\end{align}
Here, the first term is actually the classical Fisher-Rao metric. Note that the Bures metric is also called quantum Fisher information metric, which is of central role in quantum metrology.

We now apply this to BEC mixed coherent states at finite temperature. The density matrix is parameterized by $(\beta, r, \theta)$, with $\varepsilon$ as the quasi-particle energy. In these coordinates, the Bures metric becomes (Appendix \ref{app:bures})
\begin{align}\label{dsb}
  ds_B^2 &= \frac{\varepsilon^2}{8 \sinh^2(\beta\varepsilon/2)} d\beta^2 \nonumber\\
    &\quad+ \frac{1}{4}\left[1 + \frac{1}{\cosh(\beta\varepsilon)}\right](dr^2 + \sinh^2r \, d\theta^2).
\end{align}
To simplify the calculations, we introduce the coordinate transformation $\alpha = \log(\tanh(\beta \varepsilon / 4))$, which leads to
\begin{align}\label{dsb_alpha}
    ds_B^2 = \frac{1}{2} d\alpha^2 + \frac{\cosh^2 \alpha}{3 + \cosh(2\alpha)} (dr^2 + \sinh^2 r \, d\theta^2).
\end{align}
Then, we assume $\theta$ remains constant, setting $d\theta = 0$. This reduces the problem to finding the shortest path in the $(\alpha, r)$-plane. A geodesic in a Riemannian manifold, parameterized as $t \mapsto \{x^\mu\}$ with $t \in [0, 1]$, satisfies the geodesic equation \cite{nakahara_geometry_2003}
\begin{align}\label{geodesics}
    \frac{d^2 x^\gamma}{dt^2} + \Gamma_{\mu\nu}^\gamma \frac{dx^\mu}{dt} \frac{dx^\nu}{dt} = 0,
\end{align}
where $\Gamma_{\mu\nu}^\gamma = \frac{1}{2}g^{\gamma\lambda}(\partial_\mu g_{\nu\lambda} + \partial_\nu g_{\mu\lambda} - \partial_\lambda g_{\mu\nu})$ are the Christoffel symbols \cite{hetenyi_fluctuations_2023}, $g^{\gamma\lambda}$ is the inverse matrix of $g_{\gamma\lambda}$, and the Einstein summation convention is used. For our metric, these equations are
\begin{align}
  \ddot{\alpha} - \frac{2\dot{r}^2\sinh(2\alpha)}{[3 + \cosh(2\alpha)]^2} = 0, 
  \frac{d}{dt}\left[\frac{\cosh^2\alpha}{3 + \cosh(2\alpha)} \dot{r}\right] = 0. 
\end{align}

We consider the initial state $\rho(0) = \ketbra{0}{0}$, i.e. the vacuum state at zero temperature corresponding to the boundary condition $\alpha(0) = 0$ and $r(0) = 0$, and a final state $\rho(T_1, r_1)$, which implies the boundary condition $\alpha(1) = \alpha_1$ and $r(1) = r_1$. Solving the geodesic equations with details presented in Appendix \ref{app:bures}, we obtain
\begin{align}
    \alpha(t) &= -\mathrm{arcsinh}\left[ \sqrt{1 - (2c_0/c_1)^2} \sinh(c_1 t) \right], \\
    r(t) &= 2c_0 t + \mathrm{arctanh}\left[ \frac{2c_0}{c_1} \tanh(c_1 t) \right],
\end{align}
where $c_0$ and $c_1$ are constants fixed by the endpoint conditions. Along this geodesic, the metric reduces to
\begin{align}
    ds_B^2 = \left( 2c_0^2 + \frac{c_1^2}{2} \right) dt^2,
\end{align}
so the complexity is
\begin{align}
    \mathcal{C}_B = \int_0^1 \sqrt{g_{\mu\nu}^B \dot{x}^\mu \dot{x}^\nu} \, dt = \sqrt{2c_0^2 + \frac{c_1^2}{2}}.
\end{align}

\begin{figure}[h]
  \includegraphics[width=0.48\textwidth]{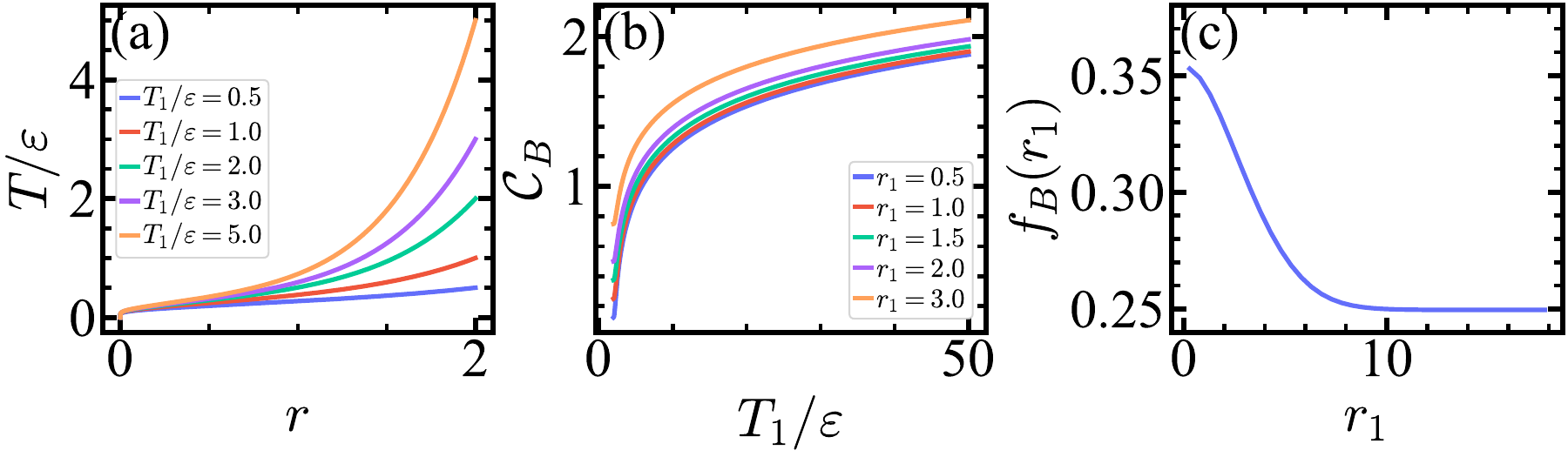}
  \caption{
    (a) Geodesics in the $(r, T)$-plane for $r(1) = 2$ and various final temperatures $T(1) = T_1$. 
    (b) Complexity $\mathcal{C}_B$ versus $T_1$ for different $r(1) = r_1$. 
    (c) Scaling factor $f_B(r_1)$ versus $r_1$.
  }
  \label{fig:bures}
\end{figure}

In Fig. \ref{fig:bures}(a), we show the geodesics in the $(r, T)$-plane for $r_1 = 2$ and various final temperatures $T(1) = T_1$. Fig. \ref{fig:bures}(b) shows the complexity versus $T_1$ for different $r_1$, revealing a scaling behavior of the complexity as a function of the final temperature for a fixed $r_1$. Through fitting data, we find that the complexity scales as 
\begin{align}
    \mathcal{C}_B \sim -f_B(r_1) \log\left[ \tanh\left( \frac{\varepsilon}{4 T_1} \right) \right],
\end{align}
where the prefactor $f_B(r_1)$, shown in Figure 1(c), decreases at small $r_1$ and saturates to approximately 0.250 for large $r_1$. This logarithmic scaling reflects how complexity grows with temperature in BEC systems.

\section{Complexity: Sjoqvist metric}\label{sec:sjoqvist}

Another generalization of the Fubini-Study metric to mixed states is the Sj\"oqvist metric \cite{sjoqvist_geometric_2000,sjoqvist_geometry_2020}. While the Bures metric relies on Uhlmann fidelity, the Sj\"oqvist metric adopts a distinct approach by define the distance between density matrices through sets of rays from the spectral decomposition of the density matrix. Below, we derive the Sj\"oqvist metric for finite-temperature BECs and investigate its implications for quantum complexity. 

We still consider the space $\mathcal{M}$ of full-rank density matrices, parameterized by coordinates $x^1, x^2, \dots, x^{\dim \mathcal{M}}$. Following Sj\"oqvist's construction \cite{sjoqvist_geometry_2020}, we consider a smooth path $t \mapsto \rho(t)$, representing the evolution of a quantum system described by density matrices. These matrices can be expressed in diagonal form as
\begin{align}
    \rho(t) = \sum_k \lambda_k(t) \ketbra{u_k(t)}{u_k(t)}.
\end{align}
The eigenvectors $\ket{u_k(t)}$ are uniquely defined up to phase factors, leading to the set of orthogonal rays
\begin{align}
    \mathcal{B}(t) = \left\{ \sqrt{\lambda_k(t)} \, e^{i f_k(t)} \ket{u_k(t)} \right\},
\end{align}
where $f_k(t) \in [0, 2\pi)$ represent phase degrees of freedom.

To define a meaningful distance between two infinitesimally close density matrices $\rho(t)$ and $\rho(t+dt)$, Sj\"oqvist proposed considering the minimal distance between their corresponding orthogonal rays $\mathcal{B}(t)=\{\sqrt{\lambda_k}\ket{u_k(t)}\}$ and $\mathcal{B}(t+dt)=\{\sqrt{\lambda_k+d\lambda_k}\ket{u_k(t+dt)}\}$

\begin{align}
  d^2(t, t + dt) =& \min_{\{f_k\}} \sum_k \bigg| \sqrt{\lambda_k(t)} \, e^{i f_k(t)} \ket{u_k(t)} \nn\\
  &- \sqrt{\lambda_k(t+dt)} \, e^{i f_k(t+dt)} \ket{u_k(t+dt)} \bigg|^2.
\end{align}
Expanding this squared norm gives
\begin{align}
    d^2(t, t + dt) =& 2 - 2 \max_{\{\phi_k\}} \sum_k \sqrt{\lambda_k(t)\lambda_k(t+dt)} \nn\\
    &\times |\braket{u_k(t)|u_k(t+dt)}| \cos \phi_k(t, t + dt),
\end{align}
where the relative phase $\phi_k(t, t + dt)$ is given by $\phi_k(t, t + dt) = \dot{f}_k(t) dt + \arg\big[1 + \langle u_k(t) | \dot{u}_k(t) \rangle dt \big] + \mathcal{O}(dt^2)$. The minimum distance condition requires $\phi_k(t, t + dt) = 0$ for all $k$, which leads to the parallel transport condition for each eigenvector:
\begin{align}
  \dot{f}_k(t) - i \langle u_k(t) | \dot{u}_k(t) \rangle = 0.
\end{align}
Expanding the Sj\"oqvist distance to second order in $dt$, we obtain the Sj\"oqvist metric \cite{sjoqvist_geometry_2020}
\begin{align}
    ds_S^2 = \sum_k \lambda_k ds_k^2 + \sum_k\frac{d\lambda_k^2}{4\lambda_k},
\end{align}
where $ds_k^2 = \bra{\dot{u}_k}(1 - \ketbra{u_k}{u_k})\ket{\dot{u}_k}dt^2$ is the Fubini-Study metric for each eigenstate $\ket{u_k}$, and $\sum_k d\lambda_k^2/4\lambda_k$ (where $d\lambda_k = \dot{\lambda}_k dt$) corresponds to the classical Fisher-Rao metric. Expressing the metric in terms of coordinate differentials $dx^\mu = \dot{x}^\mu dt$, we obtain the metric tensor $ds_S^2 = g_{\mu\nu}^S dx^\mu dx^\nu$, where
\begin{align}
  g_{\mu\nu}^S = \sum_k \left[\frac{\p_\mu\lambda_k \p_\nu\lambda_k}{4\lambda_k} + \lambda_k \braket{\p_\mu u_k|(1-\ketbra{u_k}{u_k})|\p_\nu u_k}\right].
\end{align}

Applying this metric to the BEC mixed coherent states at finite temperature, we obtain the Sj\"oqvist metric as
\begin{align}
  ds_S^2 =& \frac{\varepsilon^2}{8\sinh^2(\beta\varepsilon/2)} d\beta^2 \nn\\
  & + \frac{1}{4}(1 + \frac{1}{\cosh \beta\varepsilon - 1})(dr^2 + \sinh^2 r d^2\theta).
\end{align}
Again, we introduce the variable $\alpha=\log(\tanh(\beta\varepsilon/4))$, which leads to a compact form:
\begin{align}
  ds_S^2 = \frac{1}{2}d\alpha^2 + \frac{1+\cosh^2\alpha}{8}\left(dr^2+\sinh^2 r\, d\theta^2\right).
\end{align}

Similar to the Bures metric case, we restrict ourselves to paths with $d\theta=0$. Substituting the metric into the geodesic equation Eq.(\ref{geodesics}), we obtain
\begin{align}
  \ddot{\alpha} - \frac{\dot{r}^2\sinh(2\alpha)}{8}=0,\quad
  \frac{d}{dt}\left[\frac{1+\cosh^2\alpha}{8}\dot{r}\right]=0,
\end{align}
which can be simplified to
\begin{align}
  \dot{\alpha} = \sqrt{b_1^2 - \frac{16 b_0^2}{1+\cosh^2\alpha}},\quad 
  \dot{r} = \frac{8 b_0}{1+\cosh^2\alpha},
\end{align}
by imposing initial conditions $\alpha(0)=0, r(0)=0$ (corresponding to the vacuum reference state). The parameters $b_0$ and $b_1$ are fixed by boundary conditions at $t=1$. The final solution to the geodesic equations can be solved numerically.

Along the obtained geodesic, the Sj\"oqvist metric reduces to the simple form $ds_S^2 = b_1^2 dt^2/2$, giving the complexity
\begin{align}
  \mathcal{C}_S = \int_0^1 \sqrt{g_{\mu\nu}^S\dot{x}^\mu\dot{x}^\nu}\,dt = \frac{|b_1|}{\sqrt{2}}.
\end{align}

\begin{figure}[t]
  \includegraphics[width=0.48\textwidth]{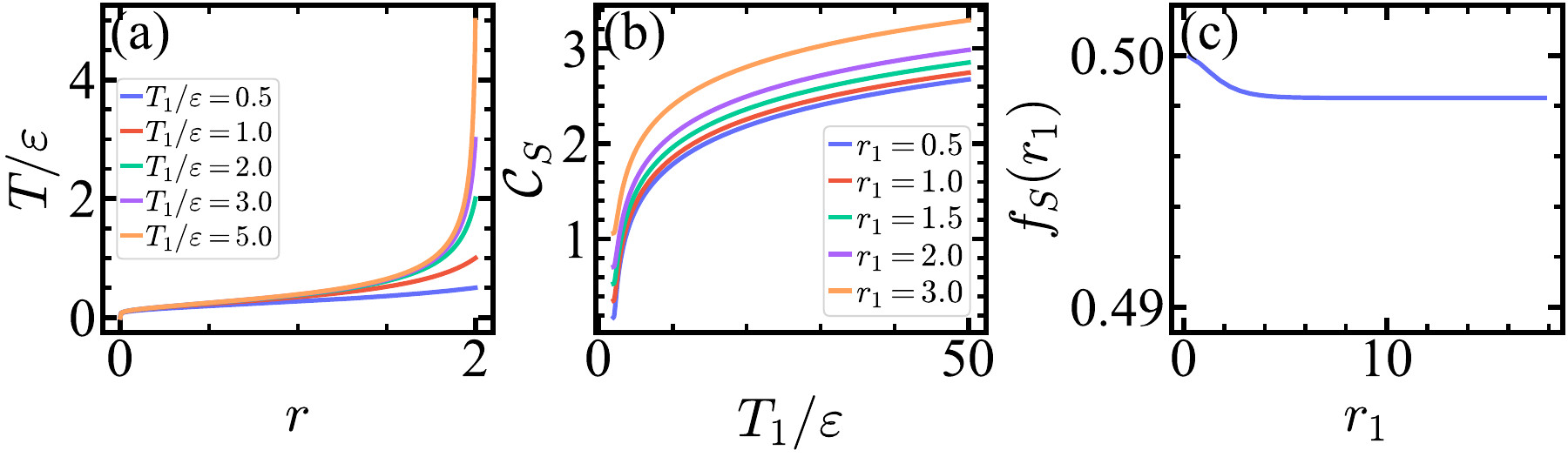}
  \caption{(a) Geodesics in the Sj\"oqvist metric for $r(1) = 2$ and varying $T(1) = T_1$. (b) Complexity $\mathcal{C}_S$ versus final temperature $T_1$ for different $r(1) = r_1$. (c) Scaling factor $f_S(r_1)$ versus $r_1$, saturating at about $0.498$.}
  \label{fig:sjoqvist}
\end{figure}

In Fig.~\ref{fig:sjoqvist}(a), we show geodesics for boundary condition $r(1) = 2$ and varying final temperatures $T(1) = T_1$ in the $(r, T)$-plane. Fig.~\ref{fig:sjoqvist}(b) shows $\mathcal{C}_S$ versus $T_1$ for different $r(1) = r_1$. Similar to the Bures metric case, a high-temperature scaling is also revealed 
\begin{align}
  \mathcal{C}_S \sim -f_S(r_1) \log\left[\tanh\left(\frac{\varepsilon}{4 T_1}\right)\right].
\end{align}
Figure~\ref{fig:sjoqvist}(c) displays the scaling factor $f_S(r_1)$, which decreases slightly with $r_1$ before saturating at approximately 0.498. As indicated by the numerical results, the complexity from the Sj\"oqvist metric is larger than that from the Bures metric, reflecting the fact that the Sj\"oqvist metric systematically assigns a larger distance than the Bures metric \cite{alsing_comparing_2023}.

\section{Nielsen complexity based on purification}\label{sec:nielsen}

Nielsen's geometric approach to quantum complexity quantifies the difficulty of transforming a reference state into a target state by identifying the minimal geodesic in the space of unitary operations \cite{nielsen_geometric_2005,nielsen_quantum_2006}. To extend this framework to mixed states, we employ purification, which embeds the mixed state into a larger Hilbert space as a pure state. In this section, we apply Nielsen's complexity to finite-temperature Bose-Einstein condensate (BEC) mixed states through purification. By systematically addressing gauge redundancies inherent to the purification process and non-uniqueness of unitary operations, we demonstrate that the Nielsen complexity, when optimized over gauge choices, coincides with the complexity derived from the Bures metric.

We begin with a continuous path of unitary operators $U(t)$ with $t\in [0, 1]$, that evolves a reference state into a target state:
\begin{align}
  U(t) = \mathcal{P} \exp\left[ -i \int_0^t dt' \sum_i y_i(t') X_i \right],
\end{align}
where $\mathcal{P}$ denotes path ordering, and the tangent vector along the path is defined as
\begin{align}
  \sum_i y_i(t) X_i = \partial_t U(t) U^\dagger(t).
\end{align}
Here, $X_i$ are Hermitian generators of the unitary group, and $y_i(t)$ are real coefficients determining the ``velocity" along the path. The complexity $\mathcal{C}(U)$ is the integral of a cost function $F(U(t), y_i(t))$ over the path:
\begin{align}
  \mathcal{C}(U) = \int_0^1 F(U(t), y_i(t)) \, dt.
\end{align}
We adopt the quadratic cost function
\begin{align}\label{cost_function}
  F(U, y_i) = \sqrt{\sum_i y_i^2},
\end{align}
which quantifies the ``speed" of traversal through the space of unitaries, resembling a geometric length.

To apply Nielsen complexity to finite-temperature BECs, we first need to construct a purification of the density matrix $\rho = \sum_\mathbf{n} \lambda_\mathbf{n} |\Psi_\mathbf{n} \rangle \langle \Psi_\mathbf{n}|$, which introduces an auxiliary Hilbert space and defines a pure state $\ket{W}$ as
\begin{align}
    \ket{W} = \sum_\mathbf{n} \sqrt{\lambda_\mathbf{n}} |\Psi_\mathbf{n}\rangle \otimes \tilde{U} |\tilde{\mathbf{n}}\rangle,
\end{align}
where $|\tilde{\mathbf{n}}\rangle = \frac{\tilde{a}_\mathbf{k}^{\dagger n_1} \tilde{a}_{-\mathbf{k}}^{\dagger n_2}}{\sqrt{n_1! n_2!}} |0\rangle$, and $\tilde{U}$ is an arbitrary unitary acting on the auxiliary system. This purification introduces a gauge freedom in the choice of $\tilde{U}$, as different choices yield equivalent density matrices upon tracing out the ancilla. Remarkably, this purified state is a coherent state of a two-component BEC and can be generated from the vacuum by a unitary operation (Appendix \ref{app:purification})
\begin{align}
    \ket{W} = U |0\rangle, \quad U = \mathcal{P} \exp\left[\int_0^1 dt \sum_i y_i(t) X_i\right],
\end{align}
where $X_i$ ($i=1,\dots,10$) are quadratic operators of creation and annihilation operators associated with the original and auxiliary modes and form the generators of the real symplectic Lie algebra $\mathfrak{sp}(4, \mathbb{R})$ \cite{wang_quantum_2024, wang_quantum_2022, penna_twospecies_2017, richaud_quantum_2017, charalambous_control_2020}. Explicitly, they can be written in the form \cite{hasebe_sp_2020}
\begin{align}
    X_i = \Psi^\dagger \kappa \gamma_i \Psi,
\end{align}
where $\Psi = (a_\mathbf{k}, \tilde{a}_{\mathbf{k}}, a_\mathbf{-k}^\dagger, \tilde{a}_{-\mathbf{k}}^\dagger)$, $\kappa = \text{diag}(1,1,-1,-1)$, and $\gamma_i$ are ten of the Dirac matrices (explicit expressions provided in Appendix~\ref{app:purification}). The generators $X_{i=1,2,3,4}$ form a $\mathfrak{u}(2)$ subalgebra.

The definition of complexity using the Nielsen approach through Purification, however, involves two gauge redundancies: (i) freedom in ancilla purification $\tilde{U}$, and (ii) non-uniqueness of unitary paths connecting vacuum to target state. Minimizing over these gauges reveals a fundamental relationship between Nielsen complexity and the Bures metric. We address these two gauge freedoms separately.

The second gauge freedom arises from the fact that unitary operations related by transformations in a subgroup that leaves the initial state invariant yield physically identical target states. Specifically, when evolving from the vacuum $\ket{0}$, there is a $U(2)$ subgroup within $Sp(4,\mathbb{R})$ leaving the vacuum invariant (similar to the $U(1)$ gauge freedom encountered in single-qubit systems with $SU(2)$ symmetry). Thus, each purified state lies in the coset space $Sp(4,\mathbb{R})/U(2)$ \cite{hasebe_sp_2020, perelomov_generalized_1986}. 

To address this gauge freedom, we consider an infinitesimal step along this path $\ket{W(t + dt)} = U(t + dt)\ket{0}$. We have
\begin{align}
  U(t + dt) &= e^{-idt\sum_i y_i(t)X_i}U(t) \nn\\
   &= U(t)U^\dag(t) e^{-idt\sum_i y_i(t)X_i}U(t) \nn\\
   &\equiv U(t)e^{-idt\sum_j \tilde{y}_j(t) X_j},
\end{align}
where the transformed coefficients are $\tilde{y}_j(t) = \sum_i y_i(t) u_{ij}$, and $u_{ij}$ arises from the adjoint action $U^\dagger(t) X_i U(t) = \sum_j u_{ij} X_j$. Using the Cartan decomposition \cite{gilmore_lie_2008}, which separates the $\mathfrak{sp}(4, \mathbb{R})$ algebra into a $\mathfrak{u}(2)$ subalgebra and its complement, we have
\begin{align}
  e^{-i dt \sum_j \tilde{y}_j X_j} = e^{-i dt \sum_{j=5}^{10} \tilde{y}_j X_j} e^{-i dt \sum_{j=1}^4 \tilde{y}_j X_j},
\end{align}
where $X_{j=1,2,3,4}$ generate $\mathfrak{u}(2)$, and $X_{j=5,\ldots,10}$ are orthogonal generators that transform the vacuum nontrivially. Since $e^{-i dt \sum_{j=1}^4 \tilde{y}_j X_j} |0\rangle = e^{i \phi} |0\rangle$, the $\mathfrak{u}(2)$ components contribute only a phase, reflecting a $\mathrm{U}(2)$ gauge freedom analogous to the $\mathrm{U}(1)$ freedom in $SU(2)/U(1)$ for spin. Thus, we can freely to tune the coefficients $\tilde{y}_{1,2,3,4}$ to minimize the cost function. For the cost function Eq.(\ref{cost_function}), we have
\begin{align}
  [F(U, y_i)]^2 dt^2 = \mathrm{Tr}(dU^\dagger dU) = \sum_j \tilde{y}_j^2 dt^2,
\end{align}
due to the orthogonality $\mathrm{Tr}(X_i X_j) = \delta_{ij}$. To minimize the cost, we set $\tilde{y}_j = 0$ for $j = 1, 2, 3, 4$, restricting the tangent vector to the directions $X_{j=5,\ldots,10}$.

For nearby states $|W(t)\rangle = U(t) |0\rangle$ and $|W(t + dt)\rangle = U(t + dt) |0\rangle$, the Fubini-Study metric is
\begin{align}
  ds_{FB}^2 &= \langle dW|dW\rangle - \langle W|dW\rangle\langle d W|W\rangle \nn\\
   &= \langle dU^\dag dU\rangle - \langle U^\dag dU\rangle\langle dU^\dag U\rangle \nn\\
   &= \sum_{ij} \tilde{y}_i \tilde{y}_j (\braket{X_i X_j} - \braket{X_i}\braket{X_j}) \nn\\
   &= \sum_{j=5}^{10} \tilde{y}_j^2 dt^2,
\end{align}
where we use the fact that $\braket{X_i X_j} - \braket{X_i}\braket{X_j}=1$ for $i=j = 5,6,...,10$, and other terms vanish. Thus, by choosing $\tilde{y}_j = 0$ for $j = 1,2,3,4$, the cost function $F(U,y_i)dt$ coincides with the Fubini-Study metric $ds_\mathrm{FS}$.

Next, we resolve the gauge freedom in $\tilde{U}$. We choose $\tilde{U}$ to maximize the Fubini-Study metric between adjacent states $|W(t)\rangle$ and $|W(t + dt)\rangle$ by imposing the parallel transport condition \cite{uhlmann_parallel_1986, uhlmann_gauge_1991, hou_local_2024}
\begin{align}
  \mathrm{Im} \langle W(t) | \frac{d}{dt} |W(t) \rangle = 0.
\end{align}
Under this condition, the overlap becomes
\begin{align}
  \langle W(t + dt) | W(t) \rangle = \mathrm{Tr} \sqrt{\sqrt{\rho(t)} \rho(t + dt) \sqrt{\rho(t)}},
\end{align}
which is the Uhlmann fidelity (See Appendix \ref{app:parallel} for more details). The Fubini-Study metric then reduces to
\begin{align}
  ds_{\mathrm{FS}}^2 &= 2 - 2 \langle W(t + dt) | W(t) \rangle \nn\\
  &= ds_B^2(\rho(t + dt), \rho(t)),
\end{align}
where $ds_B^2$ is the Bures metric between the density matrices $\rho(t)$ and $\rho(t + dt)$, obtained by tracing out the ancilla.

By minimizing the cost function over the path gauge (using Cartan decomposition) and selecting the purification gauge through parallel transport, the Nielsen complexity—computed as the integral of $F(U, y_i)$—equals the geodesic length in the Bures metric for the mixed states. This equivalence provides a consistent geometric interpretation of complexity for finite-temperature BECs, unifying circuit-based and state-based measures and validating Nielsen's framework for mixed-state systems.

\section{Conclusions}\label{sec:conclusion}

We have investigated and compared three geometric complexity measures—the Bures metric, the Sjöqvist metric, and Nielsen's circuit complexity through purification—for mixed $SU(1,1)$ coherent states in a finite-temperature BEC. Using the $SU(1,1)$ symmetry of the Bogoliubov Hamiltonian, we derived explicit expressions for these metrics on the cylindrical manifold of mixed states. Our analysis shows that both the Bures and Sjöqvist complexities increase monotonically with temperature, reflecting the growing difficulty of preparing or manipulating states as thermal excitations intensify. At high temperatures, these metrics exhibit similar scaling behaviors with the increase of final state temperature. However, the Sjöqvist metric assigns larger distances \cite{alsing_comparing_2023}, resulting in a larger scaling factor compared to the Bures metric.

A notable outcome of our analysis is the explicit equivalence we established between Nielsen complexity and the complexity defined by the Bures metric when appropriately minimized over gauge freedoms. By systematically addressing the inherent gauge freedoms involved in state purification and evolution paths, we clarified Nielsen complexity's geometric interpretation and reinforce its validity and applicability in mixed-state scenarios.

These findings have several implications. First, complexity in mixed states depends on the chosen metric, highlighting the importance of selecting an appropriate measure for specific physical applications. Second, the alignment between Bures and Nielsen complexity suggests that information-geometric methods can capture physically relevant features, enabling their use in large many-body systems where circuit optimization is impractical, such as studying quantum phase transitions through Bures geodesics \cite{zanardi_informationtheoretic_2007, carollo_geometry_2020, wang_geometric_2024}. Third, BECs provide an experimental platform to test these measures, with techniques like state tomography \cite{yi_extracting_2023} potentially estimating Bures distances and observables like the Loschmidt echo revealing complexity growth \cite{mera_dynamical_2018}.

Future directions include extending these complexity measures to other systems, such as fermionic or spin models, and investigating their behavior in non-equilibrium settings, such as quantum quenches. Additionally, developing new metrics that interpolate between the Bures and Sjöqvist approaches may offer deeper insights into mixed-state complexity.

\begin{acknowledgements}
CYW is supported by the Shuimu Tsinghua scholar program at Tsinghua University.
\end{acknowledgements}

\begin{widetext}

\appendix

\section{Bures Metric}\label{app:bures}

In this appendix, we derive the Bures metric tensor for density matrices, following the method outlined in Ref. \cite{hubner_explicit_1992}, and present a detailed calculation of the Bures metric for BEC mixed states at finite temperature, as given in Eq. (\ref{dsb}). We conclude by deriving the geodesic equations associated with this metric in the simplified $(\alpha, r)$-plane.

\subsection{General Derivation of the Bures Metric Tensor}

The Bures distance between two density matrices $\rho$ and $\rho + d\rho$ is defined as
\begin{align}
  ds_B^2(\rho, \rho + d\rho) = 2 \left[ 1 - \mathrm{Tr} \left( \sqrt{\sqrt{\rho} (\rho + d\rho) \sqrt{\rho}} \right) \right].
\end{align}
To obtain the metric tensor, we introduce a real parameter $t$ and define
\begin{align}
  A(t) = \sqrt{\sqrt{\rho} (\rho + t d\rho) \sqrt{\rho}}.
\end{align}
Expanding the Bures distance $ds_B^2(\rho, \rho + t d\rho)$ to second order in $t$, we have
\begin{align}
  ds_B^2(\rho, \rho + t d\rho) = t^2 g_{\mu\nu}^B dx^\mu dx^\nu,
\end{align}
where $g_{\mu\nu}^B$ is the Bures metric tensor. This implies
\begin{align}\label{g_a_ddot}
  g_{\mu\nu}^B dx^\mu dx^\nu = \frac{1}{2} \frac{d^2}{dt^2} ds_B^2(\rho, \rho + t d\rho) \bigg|_{t=0} = -\mathrm{Tr} \left[ \ddot{A}(t) \bigg|_{t=0} \right],
\end{align}
since the first-order term vanishes due to the normalization of $\rho$.
Then by differentiating $A(t)A(t) = \sqrt{\rho}(\rho + t d\rho)\sqrt{\rho}$ twice and set $t = 0$, we have
\begin{align}
  &\dot{A}(0)A(0) + A(0)\dot{A}(0) = \sqrt{\rho} d\rho\sqrt{\rho}, \label{a_dot}\\
  &\ddot{A}(0)A(0) + 2\dot{A}(0)\dot{A}(0) + A(0)\ddot{A}(0) = 0. \label{a_ddot}
\end{align}
Assuming $\rho$ is full-rank, we multiply Eq. (\ref{a_ddot}) from the left by $\rho^{-1}$ and take the trace:
\begin{align}\label{tr_a_ddot}
  \mathrm{Tr} \left[ \rho^{-1} \ddot{A}(0) \right] + 2 \mathrm{Tr} \left[ \rho^{-1} \dot{A}(0)^2 \right] + \mathrm{Tr} \left[ \rho^{-1} \ddot{A}(0) \rho \rho^{-1} \right] = 0,
\end{align}
which leads to
\begin{align}
  2 \mathrm{Tr} \left[ \ddot{A}(0) \right] + 2 \mathrm{Tr} \left[ \rho^{-1} \dot{A}(0)^2 \right] = 0 \ \  \Rightarrow \ \ \mathrm{Tr} \left[ \ddot{A}(0) \right] = -\mathrm{Tr} \left[ \rho^{-1} \dot{A}(0)^2 \right].
\end{align}
In the eigenbasis of $\rho = \sum_i \lambda_i \ket{u_i} \bra{u_i}$, Eq. (\ref{a_dot}) becomes
\begin{align}
  (\lambda_i + \lambda_j) \langle u_i | \dot{A}(0) | u_j \rangle = \sqrt{\lambda_i} \sqrt{\lambda_j} \langle u_i | d\rho | u_j \rangle,
\end{align}
so that
\begin{align}
  \langle u_i | \dot{A}(0) | u_j \rangle = \frac{\sqrt{\lambda_i} \sqrt{\lambda_j}}{\lambda_i + \lambda_j} \langle u_i | d\rho | u_j \rangle.
\end{align}
Substituting into Eq. (\ref{g_a_ddot}), we have
\begin{align}
  ds_B^2(\rho, \rho + t d\rho) = t^2 \tr[\rho^{-1}\dot{A}(0)^2] 
   = t^2 \sum_{ij} \frac{1}{\lambda_i} |\braket{u_i|\dot{A}(0)|u_j}|^2 
   = t^2 \sum_{ij} \frac{\lambda_j}{(\lambda_i + \lambda_j)^2} |\braket{u_i|d\rho|u_j}|^2.
\end{align}
By interchanging the indices $i$ and $j$ and setting $t = 1$, we have the Bures metric
\begin{align}
  ds_B^2(\rho, \rho + d\rho) = \frac{1}{2}\sum_{ij} \frac{|\braket{u_i|d\rho|u_j}|^2}{\lambda_i + \lambda_j}.
\end{align}
By noting the relation $d\rho = \p_\mu\rho dx^\mu$, we have the metric tensor Eq.(\ref{g_bures}).

\subsection{Bures Metric for BEC Mixed States}

For BEC mixed states at finite temperature, the density matrix is
\begin{align}
  \rho(\beta, r, \theta) = \sum_{\mathbf{n}} \lambda_{\mathbf{n}} \ket{\Psi_{\mathbf{n}}} \bra{\Psi_{\mathbf{n}}},
\end{align}
where $\lambda_{\mathbf{n}} = e^{-\beta (n_1 + n_2) \varepsilon} / (1 - e^{-\beta \varepsilon})^2$, $\mathbf{n} = (n_1, n_2)$ labels the number of quasi-particles, $\beta = 1/T$ is the inverse temperature, and $\ket{\Psi_{\mathbf{n}}}$ are coherent states defined through the displacement operator $D$ (see Eq. (\ref{dn})). The parameters $r$ and $\theta$ are the radial and angular coordinates, respectively.
The first term in the Bures metric is the classical Fisher-Rao term, which can be calculated as 
\begin{align}
  \sum_\bn \frac{(\p_\beta \lambda_\bn)^2}{4\lambda_\bn} = \frac{\varepsilon^2}{8\sinh^2(\beta\varepsilon/2)}.
\end{align}

For the second term in the Bures metric, we need to calculate the operator $D^\dag dD$.  Using the Euler decomposition of $D$ (Eq. (\ref{euler})), we have
\begin{align}
  D^\dagger dD = i d\theta \, e^{i \theta K_0} e^{i r K_2} K_0 e^{-i r K_2} e^{-i \theta K_0} - i dr \, e^{i \theta K_0} K_2 e^{-i \theta K_0} - i d\theta \, K_0.
\end{align}
Applying the $\mathfrak{u}(1,1)$ generator transformations,
\begin{align}
  e^{i r K_2} K_0 e^{-i r K_2} &= \cosh r \, K_0 - \sinh r \, K_1, \\
  e^{i \theta K_0} K_2 e^{-i \theta K_0} &= \sin \theta \, K_1 + \cos \theta \, K_2, \\
  e^{i \theta K_0} K_1 e^{-i \theta K_0} &= \cos \theta \, K_1 - \sin \theta \, K_2,
\end{align}
we have
\begin{align}
  D^\dagger \partial_\theta D &= i (\cosh r - 1) K_0 - i \sinh r (\cos \theta K_1 - \sin \theta K_2), \\
  D^\dagger \partial_r D &= -i (\sin \theta K_1 + \cos \theta K_2).
\end{align}
Defining $K_+ = K_1 + i K_2 = a_{\mathbf{k}}^\dagger a_{-\mathbf{k}}^\dagger$ and $K_- = K_1 - i K_2 = a_{\mathbf{k}} a_{-\mathbf{k}}$, we rewrite these as
\begin{align}
  D^\dagger \partial_\theta D &= i (\cosh r - 1) K_0 - \frac{i}{2} \sinh r \left( e^{i \theta} K_+ + e^{-i \theta} K_- \right), \\
  D^\dagger \partial_r D &= \frac{i}{2} \left( i e^{i \theta} K_+ - i e^{-i \theta} K_- \right).
\end{align}
Let $f_{\mathbf{m} \mathbf{n}} = \frac{(\lambda_{\mathbf{m}} - \lambda_{\mathbf{n}})^2}{\lambda_{\mathbf{m}} + \lambda_{\mathbf{n}}}$, $\gamma_\theta = \frac{\sinh r}{2} e^{i \theta}$, and $\gamma_r = \frac{i}{2} e^{i \theta}$. The second term is
\begin{align}
  &\sum_{\mathbf{m}, \mathbf{n}} f_{\mathbf{m} \mathbf{n}} \langle \Psi_{\mathbf{m}} | \partial_\mu \Psi_{\mathbf{n}} \rangle \langle \partial_\nu \Psi_{\mathbf{n}} | \Psi_{\mathbf{m}} \rangle \notag \\
  &= \sum_{\mathbf{m}, \mathbf{n}} f_{\mathbf{m} \mathbf{n}} \langle \mathbf{m} | D^\dagger \partial_\mu D | \mathbf{n} \rangle \langle \mathbf{n} | (\partial_\nu D^\dagger) D | \mathbf{m} \rangle \notag \\
  &= \sum_{\mathbf{m}, \mathbf{n}} f_{\mathbf{m} \mathbf{n}} \left[ \langle \mathbf{m} | \gamma_\mu K_+ + \gamma_\mu^* K_- | \mathbf{n} \rangle \langle \mathbf{n} | \gamma_\nu K_+ + \gamma_\nu^* K_- | \mathbf{m} \rangle \right].
\end{align}
Since $K_+$ and $K_-$ are ladder operators, non-zero terms occur only when $\mathbf{n} = \mathbf{m} \pm \mathbf{e}$ (where $\mathbf{e} = (1,1)$), and $K_0$ terms vanish due to $f_{\mathbf{m} \mathbf{m}} = 0$. Evaluating the matrix elements, we get
\begin{align}
  &= \sum_{\mathbf{m}} \left[ f_{\mathbf{m}, \mathbf{m} - \mathbf{e}} m_1 m_2 \gamma_\mu \gamma_\nu^* + f_{\mathbf{m}, \mathbf{m} + \mathbf{e}} (m_1 + 1)(m_2 + 1) \gamma_\mu^* \gamma_\nu \right] \notag \\
  &= \sum_{\mathbf{m}} f_{\mathbf{m}, \mathbf{m} + \mathbf{e}} (m_1 + 1)(m_2 + 1) (\gamma_\mu \gamma_\nu^* + \gamma_\mu^* \gamma_\nu).
\end{align}
Summing over $\mathbf{m}$, this yields
\begin{align}
  [1 + \mathrm{sech}(\beta \varepsilon)] (\gamma_\mu \gamma_\nu^* + \gamma_\mu^* \gamma_\nu).
\end{align}
Substituting $\gamma_\theta$ and $\gamma_r$, and combining with the Fisher-Rao term, the Bures metric is
\begin{align}
  ds_B^2 = \frac{\varepsilon^2}{8 \sinh^2(\beta \varepsilon / 2)} d\beta^2 + \frac{1}{4} \left[ 1 + \frac{1}{\cosh(\beta \varepsilon)} \right] (dr^2 + \sinh^2 r \, d\theta^2).
\end{align}

\subsection{Geodesic Equations for the Bures Metric}

To derive the geodesic equations, we set $d\theta = 0$ and work in the $(\alpha, r)$-plane, where $\alpha = \log(\tanh(\beta \varepsilon / 4))$. The metric becomes
\begin{align}
  ds_B^2 = \frac{1}{2} d\alpha^2 + \frac{\cosh^2 \alpha}{3 + \cosh(2\alpha)} dr^2.
\end{align}
Since the metric is diagonal, the non-zero Christoffel symbols are
\begin{align}
  \Gamma_{rr}^\alpha &= -\frac{1}{2} g^{\alpha\alpha} \partial_\alpha g_{rr} = -\frac{2 \sinh(2\alpha)}{[3 + \cosh(2\alpha)]^2}, \\
  \Gamma_{\alpha r}^r &= \Gamma_{r \alpha}^r = \frac{1}{2} g^{rr} \partial_\alpha g_{rr} = \frac{1}{2} \frac{\partial_\alpha g_{rr}}{g_{rr}},
\end{align}
where $g_{rr} = \frac{\cosh^2 \alpha}{3 + \cosh(2\alpha)}$ and $g^{\alpha\alpha} = 2$. The geodesic equations are
\begin{align}\label{geo_alpha}
  \ddot{\alpha} - \frac{2 \dot{r}^2 \sinh(2\alpha)}{[3 + \cosh(2\alpha)]^2} = 0, \ \   \ddot{r} + \frac{\dot{\alpha} \dot{r} \partial_\alpha g_{rr}}{g_{rr}} = 0.
\end{align}
Note the second equation leads to $\frac{1}{g_{rr}}\frac{d}{dt}(g_{rr}\dot{r}) = 0$, which gives
\begin{align}
  g_{rr} \dot{r} = c_0 \quad \Rightarrow \quad \dot{r} = \frac{c_0(2 + \cosh 2\alpha)}{\cosh^2\alpha} \label{r_dot},
\end{align}
where $c_0$ is a parameter that can be fixed by boundary condition. Substituting it to Eq.(\ref{geo_alpha}) gives 
\begin{align}
  \ddot{\alpha} - \frac{4c_0^2\tanh \alpha}{\cosh^2\alpha} = 0.
\end{align}
Multiplying $2\dot{\alpha}$ to both sides of above equation gives 
\begin{align}
  \frac{d}{dt}(\dot{\alpha}^2) - \frac{8c_0^2\tanh\alpha}{\cosh^2\alpha}\dot{\alpha} = 0 \ \Rightarrow \ \dot{\alpha}^2 = \int \frac{8c_0^2\tanh\alpha}{\cosh^2\alpha} d\alpha + c_1^2 = -4c_0^2\mathrm{sech}^2\alpha + c_1^2,
\end{align}
which has the solution with the boundary condition $\alpha(0) = 0$
\begin{align}
  \alpha(t) = \mathrm{arcsinh}\left[\sqrt{1 - (2c_0/c1)^2}\sinh(c_1 t)\right],
\end{align}
where $c_1$ is still a parameter fixed by the boundary condition. Substituting $\alpha(t)$ to Eq.(\ref{r_dot}) gives
\begin{align}
  \dot{r} - 2c_0\left[1 + \frac{1}{1 + [1-(2c_0/c_1)^2]\sinh^2(c_1 t)}\right] = 0,
\end{align}
which has the solution with the boundary condition $r(0) = 0$
\begin{align}
  r(t) &= 2c_0 t + \mathrm{arctanh}\left[\frac{2c_0}{c_1}\tanh(c_1 t)\right].
\end{align}
Here, the parameters $c_0$ and $c_1$ can be fixed by the boundary condition at $t = 1$. With this solution of the geodesic equations, the Bures metric takes the simple form as 
\begin{align}
  ds_B^2 = \frac{1}{2}\dot{\alpha}dt^2 + \frac{\cosh^2\alpha}{3 + \cosh(2\alpha)}\dot{r}^2 dt^2 = (2c_0^2 + c_1^2)dt^2.
\end{align}

\section{Purification of BEC mixed states and $Sp(4, \mathbb{R})$ group}\label{app:purification}
In this appendix, we will demonstrate how the single-component BEC mixed states can be purified to  two-component BEC coherent states, and review the $Sp(4, \mathbb{R})$ group \cite{hasebe_sp_2020, perelomov_generalized_1986}, which provide a powerful tool two study the two-component BEC coherent states.

\textit{Purification of BEC mixed states}--- We start from the purification of the density matrix $\rho = \sum_\bf{n} \lambda_\bf{n} \ket{\Psi_\bf{n}}\bra{\Psi_\bf{n}}$. The ancilla system we choose here is another BEC. A simple purified state is $\ket{W_0(t)} = \sum_\bf{n} \sqrt{\lambda_\bf{n}} \ket{\Psi_\bf{n}(t)} \otimes \ket{\tilde{\bn}}$, where the states $\ket{\tilde{\bn}} = \frac{1}{\sqrt{n_1!n_2!}}(\tilde{a}_{-\bk}^\dag)^{n_1}(\tilde{a}_\bk^\dag)^{n_2}|0\rangle$ live in the Fock space $\mathbb{F}_A$ of the ancilla system. However, such a purification is not unique, any unitary transformation of the ancilla system 
\begin{eqnarray}
  \ket{W} = \sum_\bf{n} \sqrt{\lambda_\bf{n}} \ket{\Psi_\bf{n}} \otimes \tilde{U}_g\ket{\tilde{\bn}}
\end{eqnarray} 
can give another purification of the density matrix. This gives us a gauge redundancy. It can be shown that the state $\ket{W}$ is actually a coherent state of two-component BEC. 

According to Eq.(\ref{dn}), $\ket{\Psi_\bn} = D\ket{\bn}$, thus, we have
\begin{align}
  \ket{W} = \sum_\bn \sqrt{\lambda_\bn} D  \ket{\bn}\otimes \tilde{U}_g \ket{\tilde{\bn}} = D\tilde{U}_g \sum_\bn \sqrt{\lambda_\bn} \ket{\bn}\otimes \ket{\tilde{\bn}} \equiv D\tilde{U}_g\ket{\psi_{2c}},
\end{align}
where we have defined $\ket{\psi_{2c}} = \sum_\bn \sqrt{\lambda_\bn} \ket{\bn}\otimes \ket{\tilde{\bn}}$. Actually, the state $\ket{\psi_{2c}}$ is nothing but a coherent state of a two-component BEC, if we view the bosonic modes in the physical system and the ones in the ancilla system as two species of bosons in the two-component BEC. More specifically, we can define the vacuum of the composite system as $\ket{0} = \ket{0}_S\otimes \ket{\tilde{0}}_A$. Then the state $\ket{\psi_{2c}}$ can be written as 
\begin{eqnarray}
  \ket{\psi_{2c}} &=& \sum_\bn \sqrt{\lambda_\bn} \frac{\ak^{\dag n_1}\amk^{\dag n_2}}{\sqrt{n_1!n_2!}} \frac{\tilde{a}_{-\bk}^{\dag n_1}\tilde{a}_{\bk}^{\dag n_2}}{\sqrt{n_1!n_2!}}\ket{0} \nn \\
   &=& \frac{1}{\sqrt{\mathcal{Z}}}\sum_\bn \frac{(e^{-\frac{\varepsilon_\bk}{2T}}\ak^\dag \tilde{a}_{-\bk}^\dag)^{n_1}}{n_1!} \frac{(e^{-\frac{\varepsilon_\bk}{2T}}\amk^\dag \tilde{a}_{\bk}^\dag)^{n_2}}{n_2!} \ket{0} \nn\\
   &=& \frac{1}{\sqrt{\mathcal{Z}}}\exp(e^{-\frac{\varepsilon_\bk}{2T}}\ak^\dag \tilde{a}_{-\bk}^\dag + e^{-\frac{\varepsilon_\bk}{2T}}\amk^\dag \tilde{a}_{\bk}^\dag) \ket{0},
\end{eqnarray}
which is a coherent state of two-component BEC. Here, we have used $\lambda_\bn = e^{-(n_1 + n_2)\varepsilon_\bk/T}/\mathcal{Z}$. Using similar technique as in Eq.(\ref{d_op}), we can rewrite $\ket{\psi_{2c}}$ as 
\begin{eqnarray}
  \ket{\psi_{2c}} = e^{i \zeta X_{10}}\ket{0},
\end{eqnarray}
where $X_{10} = -\frac{i}{2}(\ak^\dag \tilde{a}_{-\bk}^\dag + \amk^\dag \tilde{a}_\bk^\dag) + \frac{i}{2}(\ak \tilde{a}_{-\bk} + \amk \tilde{a}_\bk)$ is a generator of the $\mathfrak{sp}(4, \mathbb{R})$ Lie algebra, and $\zeta = 2\,\mathrm{arctan}(e^{-\varepsilon_\bk/2T})$.

\textit{The $\mathfrak{sp}(4,\mathbb{R})$ Lie algebra}--- The $\mathfrak{sp}(4, \mathbb{R})$ Lie algebra has ten generators, which can be written in the form 
\begin{align}
  X_i = \frac{1}{2}\Psi_\bk^\dag \kappa \gamma_i \Psi_\bk, \ \ \kappa = \mathrm{diag}(1,1,-1,-1), 
\end{align}
where $\gamma_i$'s are ten of the Dirac gamma matrices \cite{hasebe_sp_2020}
\begin{align}
  \gamma_1 &= \sigma_z\otimes I_2, \ \gamma_2 = \sigma_z\otimes\sigma_z, \ \gamma_z = \sigma_z\otimes\sigma_x, \ \gamma_4 =  I_2\otimes \sigma_y, \ \gamma_5 = i\sigma_y\otimes I_2, \nn\\
  \gamma_6 &= i\sigma_y\otimes\sigma_z, \ \gamma_7 = -i\sigma_x\otimes I_2, \ \gamma_8 = -i\sigma_x\otimes\sigma_z, \ \gamma_9 = i\sigma_y\otimes\sigma_x, \ \gamma_{10} = -i\sigma_x\otimes\sigma_x,
\end{align}
where $\sigma_{x,y,z}$ are the three Pauli matrices, and $I_2$ is the $2\times 2$ identity matrix. Here, $X_{1,2,3,4}$ form a $\mathfrak{u}(2)$ subalgebra, and the two $\mathfrak{su}(1,1)$ subalgebras are 
\begin{align}
  K_0 = (X_1 + X_2)/2, \ K_1 = (X_5 + X_6)/2, \ K_2 = (X_7 + X_8)/2,
\end{align}
which are the three generators of the $\mathfrak{su}(1,1)$ in the physical system, and
\begin{align}
  \tilde{K}_0 = (X_1 - X_2)/2, \ \tilde{K}_1 = (X_5 - X_6)/2, \ \tilde{K}_2 = (X_7 - X_8)/2,
\end{align}
which are the three generators of the $\mathfrak{su}(1,1)$ in the ancilla system. If we assume that the unitary operator $\tilde{U}_g$ can be generated by the $\mathfrak{su}(1,1)$ algebra in the ancilla system, then the purified state 
\begin{align}
  \ket{W} = D\tilde{U}_g e^{i\zeta X_9}\ket{0} \equiv U_{2c} \ket{0},
\end{align}
where $U_{2c} = D\tilde{U}_g e^{i\zeta X_{10}}$ belong to the $Sp(4, \mathbb{R})$ group. Thus, $\ket{W}$ is a coherent state of the two-component BEC.

\section{Parallel transport condition}\label{app:parallel}

In this appendix, we demonstrate that the Fubini-Study distance between nearby purified states $\ket{W(t)}$ and $\ket{W(t + dt)}$ can be minimized by appropriately choosing the gauge freedom in the purification, and that this minimal distance corresponds exactly to the Bures metric between the associated density matrices $\rho(t)$ and $\rho(t + dt)$. This minimization is achieved by imposing a parallel transport condition on the purified states.

\textit{Amplitude of the Density Matrix}--- To connect the purified states to their corresponding density matrices, we introduce the amplitude of the density matrix \cite{uhlmann_parallel_1986}. Consider a purified state in a bipartite system comprising a physical system and an ancilla system:
\begin{align}
  \ket{W} = \sum_{\mathbf{n}} \sqrt{\lambda_{\mathbf{n}}} \ket{\Psi_{\mathbf{n}}} \otimes \tilde{U}_g \ket{\tilde{\mathbf{n}}},
\end{align}
where $\ket{\Psi_{\mathbf{n}}}$ are states in the physical system, $\ket{\tilde{\mathbf{n}}}$ are Fock states in the ancilla system, $\lambda_{\mathbf{n}}$ are the eigenvalues of the density matrix $\rho = \sum_{\mathbf{n}} \lambda_{\mathbf{n}} \ket{\Psi_{\mathbf{n}}} \bra{\Psi_{\mathbf{n}}}$, and $\tilde{U}_g$ is a unitary operator acting on the ancilla.

To define the amplitude, we establish a mapping from the ancilla system to the physical system through their operators:
\begin{align}
  \tilde{a}_{\mathbf{k}} \to a_{-\mathbf{k}}, \quad \tilde{a}_{-\mathbf{k}} \to a_{\mathbf{k}},
\end{align}
and similarly for the creation operators. This mapping allows us to associate the ancilla state $\tilde{U}_g \ket{\tilde{\mathbf{n}}}$ with a corresponding state $U_g \ket{\mathbf{n}}$ in the physical system, where $U_g$ is the unitary operator induced by the mapping.

Assuming $\ket{\Psi_{\mathbf{n}}} = D \ket{\mathbf{n}}$, where $D$ is a displacement operator, we can rewrite:
\begin{align}
  U_g \ket{\mathbf{n}} = U_g D^\dagger D \ket{\mathbf{n}} = U_g D^\dagger \ket{\Psi_{\mathbf{n}}} \equiv V^T \ket{\Psi_{\mathbf{n}}},
\end{align}
defining $V^T = U_g D^\dagger$. The amplitude $W$ of the density matrix is then:
\begin{align}
  W = \sum_{\mathbf{n}} \sqrt{\lambda_{\mathbf{n}}} \ket{\Psi_{\mathbf{n}}} \bra{\Psi_{\mathbf{n}}} V = \sqrt{\rho} V,
\end{align}
where $V$ is the transpose of $V^T$, and $\sqrt{\rho} = \sum_{\mathbf{n}} \sqrt{\lambda_{\mathbf{n}}} \ket{\Psi_{\mathbf{n}}} \bra{\Psi_{\mathbf{n}}}$ is the square root of the density matrix. This amplitude encapsulates the gauge freedom introduced by the unitary $V$.

\textit{Minimizing the Fubini-Study Distance}--- The Fubini-Study distance between two purified states $\ket{W_1}$ and $\ket{W_2}$ is defined as:
\begin{align}
  ds_{\text{FS}}^2(\ket{W_1}, \ket{W_2}) = 2 - 2 \left| \braket{W_1 | W_2} \right|^2.
\end{align}
Since $\braket{W_1 | W_2} = \mathrm{Tr}(W_1^\dagger W_2)$ \cite{hubner_computation_1993}, and the purification involves gauge freedom through $U_{g,1}$ and $U_{g,2}$ (or equivalently $V_1$ and $V_2$), we seek to minimize the distance by maximizing the overlap:
\begin{align}
  ds_{\text{FS,min}}^2 = 2 - 2 \max_{V_1, V_2} \left| \mathrm{Tr}(W_1^\dagger W_2) \right|^2 = 2 - 2 \max_{V_1, V_2} \left| \mathrm{Tr}(\sqrt{\rho_1} V_1^\dagger \sqrt{\rho_2} V_2) \right|^2.
\end{align}
Define $A = \sqrt{\rho_1} V_1^\dagger \sqrt{\rho_2} V_2$. Using polar decomposition, $A = |A| U_A$, where $|A| = \sqrt{A A^\dagger}$ is positive semi-definite and $U_A$ is unitary. To maximize $|\mathrm{Tr}(A)|$, apply the Cauchy-Schwarz inequality:
\begin{align}
  |\mathrm{Tr}(A)| = |\mathrm{Tr}(\sqrt{|A|} \sqrt{|A|} U_A)| \leq \sqrt{\mathrm{Tr}(|A|) \mathrm{Tr}(U_A^\dagger |A| U_A)} = \mathrm{Tr}(|A|),
\end{align}
with equality when $\sqrt{|A|} U_A = \sqrt{|A|}$, implying $U_A = I$. Thus, $A = |A|$, and since $|A|$ is positive, $A$ must be Hermitian and positive definite:
\begin{align}
  A = A^\dagger > 0.
\end{align}
This condition translates to:
\begin{align}
  W_1^\dagger W_2 = V_1^\dagger \sqrt{\rho_1} \sqrt{\rho_2} V_2 = V_2^\dagger \sqrt{\rho_2} \sqrt{\rho_1} V_1 = W_2^\dagger W_1 > 0.
\end{align}
For a continuous path $\ket{W(t)}$, consider $W_1 = W(t)$ and $W_2 = W(t + dt)$. The condition becomes:
\begin{align}
  W^\dagger(t) \dot{W}(t) = \dot{W}^\dagger(t) W(t) > 0,
\end{align}
where $\dot{W}(t) = \frac{d}{dt} W(t)$. This implies the parallel transport condition:
\begin{align}
  \mathrm{Im} \braket{W(t) | \frac{d}{dt} | W(t)} = \mathrm{Tr}(W^\dagger \dot{W} - \dot{W}^\dagger W) = 0,
\end{align}
ensuring the imaginary part vanishes when the real part is maximized.

\textit{Connection to the Bures Metric}--- When the parallel transport condition holds, the maximal overlap is $\mathrm{Tr}(|A|)$, and the minimal Fubini-Study distance between $\ket{W(t)}$ and $\ket{W(t + dt)}$ is:
\begin{align}
  ds_{\text{FS,min}}^2(t, t + dt) = 2 - 2 \mathrm{Tr}(|A|) = 2 - 2 \mathrm{Tr} \sqrt{\sqrt{\rho(t)} \rho(t + dt) \sqrt{\rho(t)}}.
\end{align}
This expression matches the Bures metric $ds_B^2(\rho(t), \rho(t + dt))$, defined as the infinitesimal distance between density matrices. Thus, by selecting the gauge to satisfy the parallel transport condition, the minimal Fubini-Study metric on the purified states coincides with the Bures metric on the corresponding density matrices.

\end{widetext}

\bibliography{ref.bib}

\end{document}